\documentclass{aa}
\usepackage{graphicx}
\usepackage{epsfig,graphicx,rotating,portland,fancyheadings,longtable}
\usepackage{multirow}
\usepackage{txfonts}

\begin{document}
\title{GRB\,100614A and GRB\,100615A: two extremely dark GRBs}

\author{
V. D'Elia and G. Stratta
}

\institute
{
ASI-Science Data Center, Via Galileo Galilei, I-00044 Frascati, Italy\\
INAF-Osservatorio Astronomico di Roma, Via Frascati 33, I-00040 Monteporzio Catone, Italy\\
}

  \abstract 
  { Dark gamma-ray bursts (GRBs) are sources with a low
    optical-to-X-ray flux ratio. Proposed explanations for this
    darkness are: i) the GRB is at high redshift ii) dust in the GRB
    host galaxy absorbs the optical/NIR flux iii) GRBs have an
    intrinsically faint afterglow emission.}
  { We study two dark GRBs discovered by {\it Swift}, namely,
    GRB\,100614A and GRB\,100615A. These sources are bright in the
    X-rays, but no optical/NIR afterglow has been detected for either
    source, despite the efforts of several follow-up campaigns that
    have been performed since soon after the GRB explosion.}
  {We analyze the X-ray data and collect all the optical/NIR upper
    limits in literature for these bursts. We then build
    optical-to-X-ray spectral energy distributions (SEDs) at the times
    at which the reddest upper limits are available, and we model our
    SEDs with the attenuation curves of the Milky Way (MW), Small
    Magellanic Cloud (SMC), and one obtained for a sample of starburst
    galaxies. }
  {We find that to explain the deepest NIR upper limits assuming
    either a MW or SMC extinction law, the visual extinction towards
    GRB\,100614A is $A_V>47$ mag, while for GRB\,100615A we obtain
    $A_V>58$ mag using data taken within one day after the burst and
    $A_V>22$ mag even 9.2 days after the trigger.  }
  { A possible explanation to these unlikely values is that optical
    radiation and X-rays are not part of the same synchrotron
    spectrum. An alternative, or complementary explanation of the
    previous possibility, involves greyer extinction laws. A starburst
    attenuation curve gives $A_V>11$ ($A_V>15$) for GRB\,100614A
    (GRB\,100615A) before 1 day after the burst, which is less
    extreme, despite still very high. Assuming high redshift in
    addition to extinction, implies that $A_V>10$ at $z=2$ and
    $A_V>4-5$ at $z=5$, regardless of the adopted extinction
    recipe. These lower limits are well above the $A_V$ computed for
    previous GRBs at known redshift, but not unlikely. A different,
    exotic possibility would be an extremely high redshift origin
    ($z>17$ given the missing K detections). PopIII stars are expected
    to emerge at $z \sim 20$ and can produce GRBs with energies well
    above those inferred for our GRBs at these redshifts. However,
    high $N_H$ values (above the Galactic ones) towards our GRBs
    challenge this scenario. Mid- and far-IR late afterglow ($>10^5$s
    after trigger) observations of these extreme class of GRBs can
    help us to differentiate between the proposed scenarios.}

  \authorrunning {} 
\titlerunning {GRB\,100614A and 100615A: two
    peculiar dark GRBs}

\maketitle
%

\section{Introduction}

Long-duration gamma-ray bursts (GRBs) are high energy phenomena linked
to the death of massive stars, emitting most of their radiation in the
hundreds of keV range.  The gamma-ray (or prompt) event is followed by
an afterglow at longer wavelengths, which is crucial to understand the
physics of these sources and investigate the nature of their
surrounding medium.

While the X-ray (0.1-10 keV) afterglow is virtually detected for all
GRBs, the optical/near infrared (NIR) one is more elusive. The first
optical afterglow detection (van Paradijs et al. 1997) suggested the
idea that every GRB had an optical counterpart.  However, just a few
months later, the search for the optical afterglow associated with
GRB\,970828 was unsuccessful (Groot et al. 1998), leading to the
definition of `dark burst' to underline a GRB with an X-ray
counterpart, but not an optical one. GRB\,970828 was not an isolated
event, and the low success rate in the optical/NIR detection of
afterglows became a hot GRB topic (see e.g., Fynbo et al. 2001 and
Lazzati et al. 2002). Initially, this lack of detection was widely
associated with the delay time between the GRB and the optical
observations, since ground-based facilities could be on target only
several hours after the trigger, when the afterglow faded below their
sensitivity threshold.

The main scientific driver of the {\it Swift} satellite (Gehrels et
al. 2004) was to facilitate the GRB afterglow detection through a
quick repointing with its narrow field instruments (XRT in the 0.3-10
keV band, Burrows et al. 2005 and UVOT at UV/optical wavelengths,
Roming et al. 2005) and a fast dissemination of the GRB coordinates
worldwide. Despite these instruments repoint the target fewer than two
minutes after the prompt event, the UVOT detection rate is just $\sim
40\%$ (Roming \& Mason 2006). However, the quick detection with XRT
and the increasing number of ground-based automated facilities
dramatically improved the optical follow-up success, reducing the
fraction of dark GRBs.

The optical darkness can be ascribed to different factors (see e.g.,
Fynbo et al. 2001, Perley et al. 2009).  First, the GRB can be at high
redshift, so that the Lyman $\alpha$ absorption prevents optical
identifications. Second, dust in the GRB host galaxy or along the line
of sight can absorb the optical afterglow counterpart. Finally, the
optical faintness can have an intrinsic origin.

The definition of dark GRB evolved as more data became available.
Originally, a dark GRB was an event with an X-ray afterglow but no
optical detection (Fynbo et al. 2001). Then, brightness and time
limits were added to make this definition more specific, e.g., $R>23$
within $12$ hr after the prompt event. Finally, the physics of the GRB
was involved in the dark/bright dichotomy. The basic prediction of the
fireball model (M\'esz\'aros \& Rees 1997), which is commonly invoked
to explain the afterglow, is that synchrotron radiation is responsible
for the optical-to-X-ray emission. According to this model, the
spectral index in the optical and X-rays is a function of the
power-law index $p$ of the electron energy distribution, and is
$(p-1)/2$ or $p/2$ depending where the synchrotron cooling (injection)
frequency in the slow (fast) cooling regime is located with respect to
the observed band (Sari et al. 1998). The simplest blast wave models
(Wijers et al. 1997; Sari et al. 1998) predict $p \ge 2$, which
translates to a spectral index $\ge 0.5$.  In this framework,
Jakobsson et al. (2004) proposed a classification in which dark GRBs
had an optical-to-X-ray spectral index $\beta_{OX}<0.5$. Despite many
GRBs behaving this way, it has been shown that $p<2$ is achievable
both observationally and theoretically, by introducing a high-energy
cut-off in the electron distribution (see e.g., Dai \& Cheng 2001,
Starling et al. 2008 and references therein). Thus, $\beta_{OX}$ can
in principle be lower than $0.5$ in the synchrotron model. Rol et
al. (2005) proposed a classification criterion based on the faintest
allowed extrapolation of the X-ray flux to the optical regime,
inferring the $p$ value from both the X-ray spectral and temporal
indices according to the fireball closure relationships (see e.g. Sari
et al. 1998).

Another approach was proposed by van der Horst et al. (2009).
They noted that, regardless of many assumptions about the specific
electron energy distribution, if both the optical and X-ray radiation
are produced by synchrotron emission from the same source, the
spectral indices in the optical ($\beta_O$) and X-ray ($\beta_X$)
bands are linked. In particular, $\beta_O=\beta_X -0.5$ if the cooling
frequency lies between the optical and the X-rays, and
$\beta_O=\beta_X$ otherwise. Thus, the optical-to-X-ray spectral index
allowed range is $\beta_X-0.5\le\beta_{OX}\le\beta_X$, with
$\beta_{OX}=\beta_X-0.5$ if a spectral break is present just below the
lowest X-ray energy detected. GRB afterglows with
$\beta_{OX}<\beta_X-0.5$ are classified as dark in this picture.

The complex nature of the X-ray lightcurve revealed by {\it Swift}
between a few tens of seconds and several hours after the trigger,
places some constraints on the dark GRB identification methods
mentioned above based on the assumption that both the optical and
X-ray emission have the same origin, that is, lie on the same energy
spectrum.  The steep initial decay observed in most X-ray afterglow
light curves and not at optical wavelengths, is indeed thought to be
produced by the prompt emission, thus by a mechanism different from
the one responsible for the afterglow.  The optical and X-ray spectral
energy distributions should then be compared after the initial steep
decay in order to be sure that the prompt emission is not dominating
the X-ray afterglow emission and that early optical flashes from
reverse shocks are not present (see, e.g., van der Horst et al. 2009).
In addition, in about $70\%$ of Swift/XRT detected afterglows, the
X-ray light curve between a few hundreds of seconds (thus at the end
of the steep decay) up to several minutes/hours shows a plateau phase
that is on average not present in the optical counterpart (e.g. Nousek
et al. 2006, Liang et al. 2009).  This behavior is not predicted by
the standard fireball model. After the plateau phase, alternatively
called the "shallow phase", the X-ray light curve decays following the
afterglow behavior expected from the fireball model. Interestingly,
the transition from the "shallow phase" to the "normal phase" is not
accompanied by any spectral variation (see e.g. Liang et
al. 2009). The interpretation of the so-called "shallow phase"
observed in the Swift GRB X-ray afterglow remains unclear. Therefore,
the spectral extrapolation to the optical observation times should be
interpreted with caution.

Here we study the `darkness' properties of two GRBs (namely,
GRB\,100614A and GRB\,100615A), which are very bright in X-rays, but
are not detected in the optical/near-IR band and have no reported host
galaxy candidate. The paper is organized as
follows. Sect. 2. summarizes the discovery and observations of these
two sources, Sect. 3 presents instead our analysis method, Sect. 4
illustrates our results, and in Sect. 5 we discuss our findings and
draw our conclusions. In the following, we assume a concordance
cosmology with $H_0=70$ km s$^{-1}$ Mpc$^{-1}$, $\Omega_{\rm m} =
0.3$, $\Omega_\Lambda = 0.7$. Decay, photon, and spectral indices are
indicated with $\alpha$, $\Gamma$, and $\beta$, following the standard
convention $t^{-\alpha}$, $N_{ph}^{-\Gamma}$, and $\nu^{-\beta}$,
respectively.

\begin{table*}[ht]
\begin{center}
  \caption{GRB\,100614A and GRB\,100615A ground-based observation
    summary. 
}
{\footnotesize
\smallskip
\begin{tabular}{|l|c|c|c|c|c|c|c|c|c|c|}
\hline 
Source     & $t-T$ & flux density & U & G      & R       & I      & Z    &  J      & H       & K        \\
     & min & $\mu$Jy at $1.7$ keV  & mag & mag      & mag       & mag     & mag    &  mag      & mag       & mag         \\
\hline
GRB\,100614A$^a$ & $6.9$    &     29.6 & -           & -       & $>18$   & -       & -       & -       & -       & -        \\
GRB\,100614A$^b$ & $29$     & 1.26      & -           & -       & $>21$   & -       & -       & -       & -       & -        \\
GRB\,100614A$^c$ & $31$     &   1.08   & -           & -       & -       & $>21$   & -       & -       & -       & -        \\
GRB\,100614A$^d$ & $80$ & 1.20     & -           & -       & $>24.0$ & $>22.8$ & -       & -       & -       & -        \\
GRB\,100614A$^e$ & $60$     &    0.25  & -           & -       & -       & $>24$   & -       & -       & -       & -        \\
{\bf GRB\,100614A$^f$} & {\bf 258}  & {\bf 0.10} & {\bf $>$25.2}&{\bf$>$27.1}  & {\bf$>$26.4} & {\bf$>$25.9} & {\bf$>$24.9} & -       & -       & -        \\
{\bf GRB\,100614A$^g$} & {\bf 570  } & {\bf 0.07} & -           & -       & -       & -       & -       & {\bf $>$22.7} & -       & {\bf $>$21.6 } \\
GRB\,100614A$^h$ & $1330$   & 0.05       & -           & -       & $>21.9$ & -       & -       & -       & -       & -        \\
\hline
{\bf GRB\,100615A$^{i,AB}$} & {\bf 24 }    & {\bf 9.84}  & -           & {\bf $>$24.2} & {\bf $>$23.9} & {\bf $>$22.9} & {\bf $>$22.5} & {\bf $>$21.4} & {\bf $>$20.7} & {\bf $>$20.3}  \\
GRB\,100615A$^j$ & $28$     & $9.69$  & -           & -       & -       & $>24.0$ & -       & -       & -       & -        \\
GRB\,100615A$^k$ & $114$    & $4.92$  & -           & -       & -       & -       & -       & $>18.8$ & $>17.9$ & $>16.6$  \\
{\bf GRB\,100615A$^l$} & {\bf 330}    & {\bf 1.97}  & -           & -       & -       & -       & -       & -       & -       & {\bf $>$20.9} \\
{\bf GRB\,100615A$^m$} & {\bf 13200}  & {\bf 0.013} & -           & -       & -       & -       & -       & -       & -       & {\bf $>$22.2}  \\
\hline
\end{tabular}

}
\end{center}

$^a$: BOOTES-2 TELMA 0.6m robotic telescope data from Jelinek et al. (2010); 
$^b$: 2-m Liverpool Telescope data from Mundell et al. (2010); 
$^c$: 2-m Liverpool Telescope data from Mundell et al. (2010); 
$^d$: NOT data from Malesani et al. (2010)
$^e$: 4.2m William Herschel Telescope data from Levan et al. (2010)
$^f$: GTC 10.4-m data from Guziy et al. (2010); 
$^g$: Gemini-North data from Cenko et al. 2010b; 
$^h$: AZT-11 telescope data from Shakhovskoy et al. (2010); 
$^i$: GROND data from Nicuesa et al. (2010); 
$^j$: NTT/Ultracam data from Dhillon et al. (2010); 
$^k$: PAIRITEL data from Morgan et al. (2010); 
$^l$: Gemini-North data from Cenko et al. (2010a); 
$^m$: Gemini-North data from Perley et al. (2010);

$^{AB}$: Magnitudes given in the AB system 
\end{table*}

\section{The dataset}

\subsection{GRB\,100614A}

GRB\,100614A was discovered by {\it Swift} on June 14, 2010, at
21:38:26 UT (Stratta et al. 2010). The Swift/BAT (15-350 keV) coded
mask-weighted light curve shows a relatively smooth peak starting at
$\sim T-10$ s, peaking around $T+50$ s, and ending at $\sim T+275$ s,
where T is the trigger time. The burst duration has been estimated as
$T_{90}=225\pm55$ s, thus classifying this burst among the long
GRBs. The time-averaged spectrum is best fit by a simple power-law
model with photon index $1.88\pm0.15$. The 15-150 keV fluence is
$(2.7\pm0.2)\times 10^{-6}$ erg cm$^{-2}$ (Sakamoto et al. 2010).

XRT and UVOT observations were initiated 133 s and 873 s after the BAT
trigger, respectively.  XRT immediately found a bright afterglow. The
astrometrically corrected X-ray afterglow position (using the XRT-UVOT
alignment and matching UVOT field sources to the USNO-B1 catalogue)
with 3.4 ks of exposure, is RA(J2000)=17h 33m 59.82s and
Dec(J2000)=+49d 14$'$ 03.6$''$ ($1.7''$ radius, 90\% confidence,
Osborne et al. 2010), while UVOT did not detect the optical
counterpart despite the low Galactic reddening in the direction of
this burst (E(B-V) of 0.03 mag, Schlegel et al. 1998). The XRT
follow-up (Margutti et al. 2010) continued up to about eight days
after the trigger and the overall light curve shows the canonical
"steep-flat-normal" decay (e.g. Nousek et al. 2006).

XMM-Newton also observed GRB\,100614A, starting at 04:12 UT, on June
15, 2010, for an exposure of 42 ksec (Schartel 2010), but no results
from data analysis have yet been published. Several optical follow-up
observations have been performed with no optical/NIR counterpart
detection. Table 1 summarizes the optical/NIR upper limits to the
GRB\,100614A afterglow, together with its X-ray flux at $1.7$ keV at
the corresponding acquisition times. Magnitudes are in the Vega system
(unless otherwise stated, see footnotes), but not corrected for
Galactic extinction. The mean observational epochs from trigger are
quoted in the second column. Bold face characters indicate the epochs
with the reddest and deepest upper limits. Fig. 1 (top panel) shows
the X-ray light curve and all the optical/NIR upper limits reported in
Table 1.

\begin{figure}
\centering
\includegraphics[angle=-0,width=9cm]{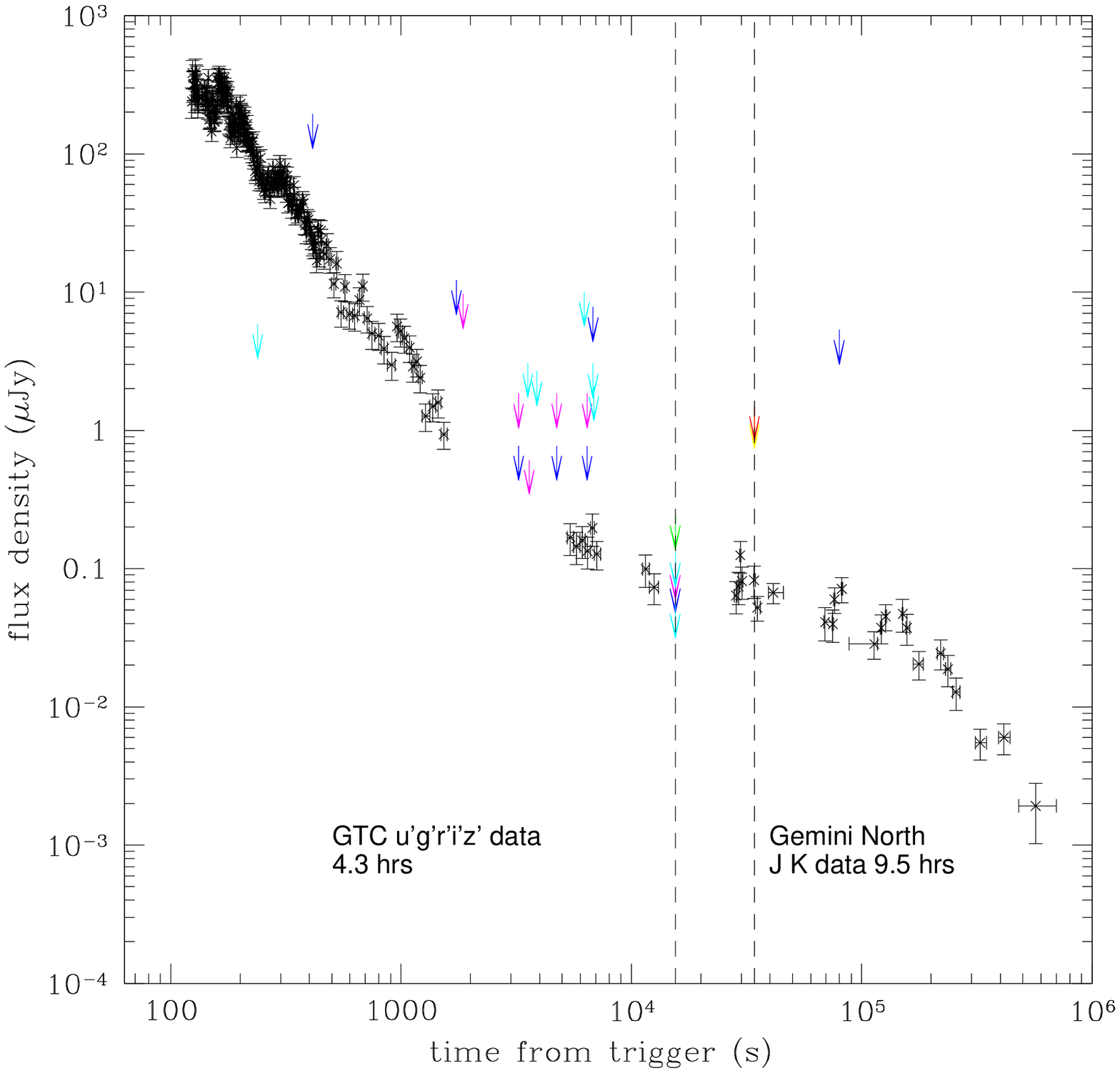}
\includegraphics[angle=-0,width=9cm]{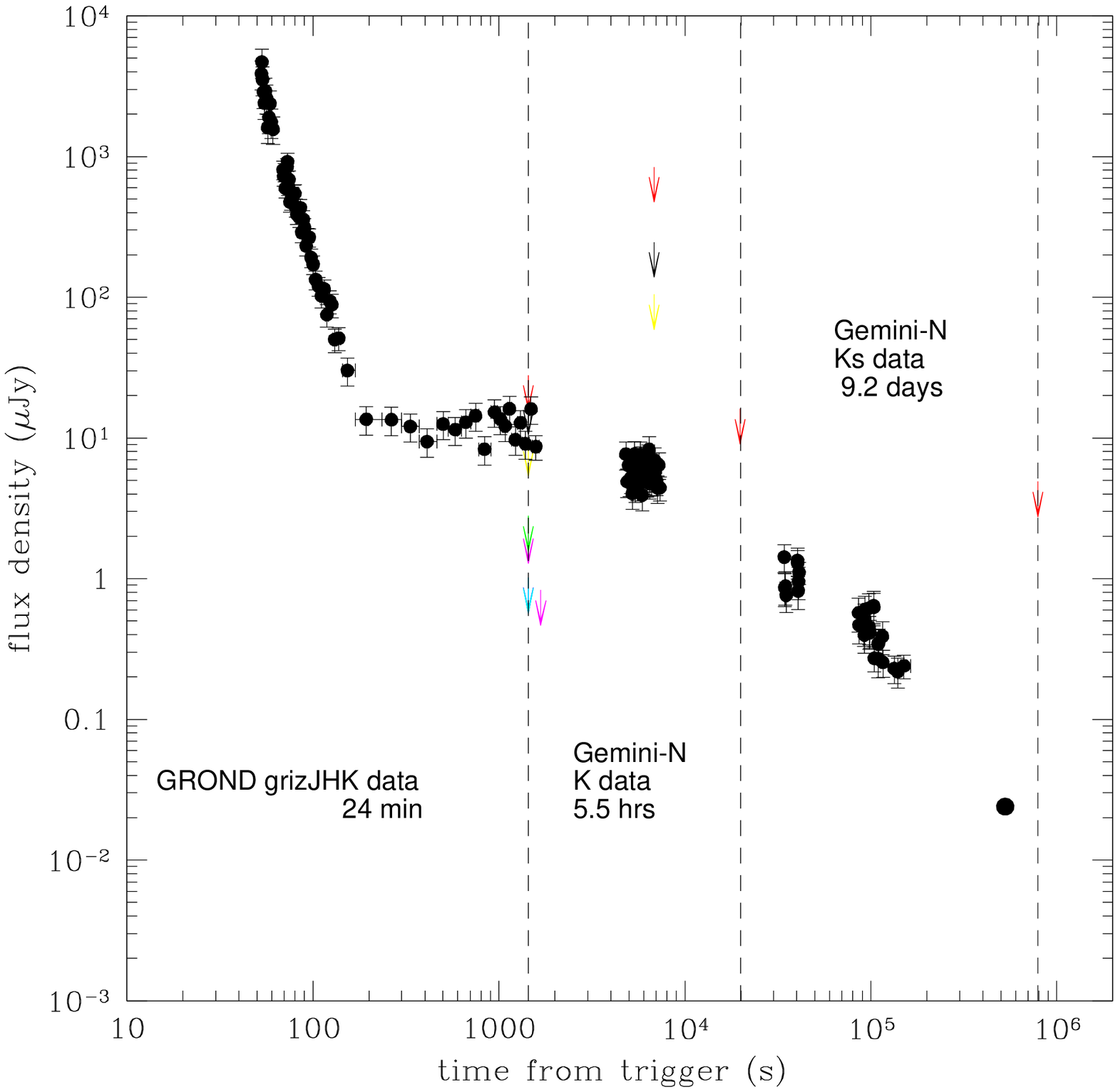}
\caption{The GRB\,100614A (upper panel) and GRB\,100615A (bottom
  panel) X-ray lightcurves, together with all the optical/NIR upper
  limits estimated using several ground-based facilities (see Table 1
  for details). Cyan, blue, magenta, green, yellow, black and red
  upper limits represents the g' (and all those bands bluer than g'),
  R, I, z, J, H, K bands, respectively. Vertical dashed lines
  emphasize the reddest observations and the time at which they have
  been obtained. Such observations are used in our analysis. The
  faintest X-ray data in the bottom panel draws the GRB\,100615A {\it
    Chandra} observation (Butler et al. 2010).}
\label{spe1}
\end{figure}

\subsection{GRB\,100615A}

GRB\,100615A was discovered by {\it Swift} on June 15, 2010, at
01:59:03 UT (D'Elia et al. 2010). The Swift/BAT (15-350 keV) coded
mask-weighted light curve shows three slightly overlapping FRED (fast
rise exponential decay) peaks. The burst duration has been estimated
as $T_{90}=39\pm2$ s. The time-averaged spectrum is best fit by a
simple power-law model with photon index of $1.87\pm0.04$.  The
fluence in the 15-150 keV band is $(5.0\pm0.1)\times 10^{-6}$ erg
cm$^{-2}$.

XRT and UVOT observations were initiated about a minute after the BAT
trigger. XRT immediately found a bright afterglow. The astrometrically
corrected X-ray afterglow position (using the XRT-UVOT alignment and
matching UVOT field sources to the USNO-B1 catalogue) with 5.2 ks of
exposure is RA(J2000) = 11h 48m 49.26s and Dec(J2000) = -19d 28'
52.4$''$ ($1.4''$ error radius at the 90\% confidence, Osborne et
al. 2010), while UVOT did not detect the optical counterpart despite
the low Galactic reddening in the direction of this burst (E(B-V) of
0.05 mag, Schlegel et al. 1998). XRT follow-up continued up to about
two days after the trigger.  As for GRB\,100614A, the overall light
curve shows the canonical "steep-flat-normal" decay.

A DDT was issued and approved to observe the X-ray afterglow with {\it
  Chandra} (Butler et al. 2010). The source was still visible six days
after the burst, and the position was refined to RA, Dec (J2000) =
11$^h$ 48$^m$ 49$^s$.34, -19d 28' 52.0" with uncertainty of $0.6"$,
and the spectral parameters were found to be in agreement with those
obtained analyzing the XRT data (see next section). The optical/NIR
afterglow was searched starting from a few minutes from the BAT
trigger, using four ground-based facilities, but only upper limits to
its emission could be set. Table 1 summarizes the optical/NIR upper
limits for the GRB\,100615A afterglow, together with its X-ray flux at
$1.7$ keV at the corresponding acquisition times. Magnitudes are in
Vega system (unless otherwise stated, see footnotes), but not
corrected for Galactic extinction. The mean observational epochs from
trigger are quoted in the second column. Bold face characters indicate
the epochs with the reddest and deepest upper limits.  Fig. 1 (bottom
panel) shows the X-ray light curve and all the optical/NIR upper
limits for GRB\,100615A.

\section{Data analysis}

We extract the optical to X-ray spectral energy distribution (SED) for
both GRBs by selecting those epochs at which we have the deepest and
reddest observations (i.e. less affected by any dust extinction), to
constrain at best the intrinsic optical afterglow flux upper limit.
In addition, we attempt to select an epoch not too close to the
initial X-ray steep decay, which is thought to be produced by a
different component than the one responsible for the afterglow
emission. Magnitudes were corrected for Galactic absorption.

Swift/XRT data were calibrated, filtered and screened using the XRTDAS
package included in the HEAsoft distribution (v6.10) as described in
the XRT Software User's
Guide\footnote{http://swift.gsfc.nasa.gov/docs/swift/analysis/}.
Unabsorbed X-ray fluxes were estimated at the selected epochs (see
below).

A broken power-law model was fitted to the data following the van der
Horst et al. (2009) method, therefore fixing the SED normalization and
the high-energy spectral index to the value obtained from our X-ray
data analysis (within its $90\%$ confidence range), the spectral break
at the X-ray energies and the optical to X-ray spectral index as
$\beta_{OX}=\beta_X-0.5$.

We model the optical suppression from the X-ray extrapolation assuming
either a Milky Way (MW) or a Small Magellanic Cloud (SMC) extinction
curve.  We also test the attenuation curve obtained for a sample of
starburst galaxies (Calzetti et al. 1994). We consider the upper
limits as positive detections, and we verify that the model-predicted
fluxes are consistent with the data, i.e. equal to or below the upper
limits.

\subsection{GRB\,100614A}
For this burst, the X-ray steep decay phase ends about 30 minutes
after the trigger, while the plateau stops 2.27 days after the
trigger.  The reddest and deepest upper limits are those obtained 258
and 570 minutes after the trigger with the 10.4-m Gran Telescopio
Canarias (GTC) in the $ugriz$ photometric bands (Guizy et al. 2010)
and with Gemini Near InfraRed Imager on the 8-m Gemini North telescope
in the J and K bands (Cenko et al. 2010b), respectively (see Table
1). The data of the first observation were photometrically calibrated
using the SDSS stars and the upper limits are at the $3 \sigma$
confidence level (Guziy et al. 2010; Guziy, private
communication). The data of the second observation were calibrated
using three 2MASS stars and are given at the $5 \sigma$ confidence
level. The RMS spread evaluated by comparing these data with
calibrators is $0.1$ ($0.2) $ mag in the J (K) band (Cenko et
al. 2010b; Cenko, private communication). All the available magnitudes
were corrected for Galactic reddening in the direction of this burst.
The optical upper limits corresponding to the magnitudes of the first
observation are the following (from the U to the z band): $f_{U}<0.13
,f_{g}<0.056$, $f_{r}<0.085$, $f_{i}<0.11$, $f_{z}<0.24$ $\mu$Jy. The
optical upper limits corresponding to the J and K data points of the
second observation are $f_J<1.3$ $\mu$Jy and $f_K<1.5$ $\mu$Jy,
respectively.

The 0.3-10 keV energy spectrum integrated between 10 ks and 100 ks
after the trigger, that is during the plateau phase, is closely fitted
by an absorbed power law with an equivalent hydrogen column density
$N_H=(1.6\pm0.7)\times10^{21}$ cm$^{-2}$ beyond the Galactic one
measured at the position of the X-ray afterglow
($N_{H,Gal}=2.2\times10^{20}$ cm$^{-2}$, Kalberla et al. 2005) and a
photon index of $\Gamma=2.5\pm0.3$ ($90\%$ confidence range). The
count rate to unabsorbed flux conversion factor is $6\times10^{-11}$
erg cm$^{-2}$ cts$^{-1}$.

The X-ray spectrum does not vary significantly during the plateau
phase, thus we assume this spectral shape for the epochs at which we
wish to extract the broad-band SEDs, and we compute the X-ray
normalization level from the light curve at the given epochs. The
count rate 258 and 570 minutes after the trigger is $c_1=0.03$ counts
s$^{-1}$ and $c_2=0.02$ counts s$^{-1}$. Converting these values into
unabsorbed flux using the conversion estimated factor, we find
$f_1=1.8\times10^{-12}$ erg cm$^{-2}$ s$^{-1}$ and
$f_2=1.2\times10^{-12}$ erg cm$^{-2}$ s$^{-1}$. The flux densities at
1.7 keV (logarithmic mean of the XRT energy range) are therefore
$f_{\nu_1}=0.10$ and $f_{\nu_2}=0.07$ $\mu$Jy 258 and 579 minutes
after the trigger, respectively.

\begin{figure}
\centering
\includegraphics[angle=-0,width=4.4cm]{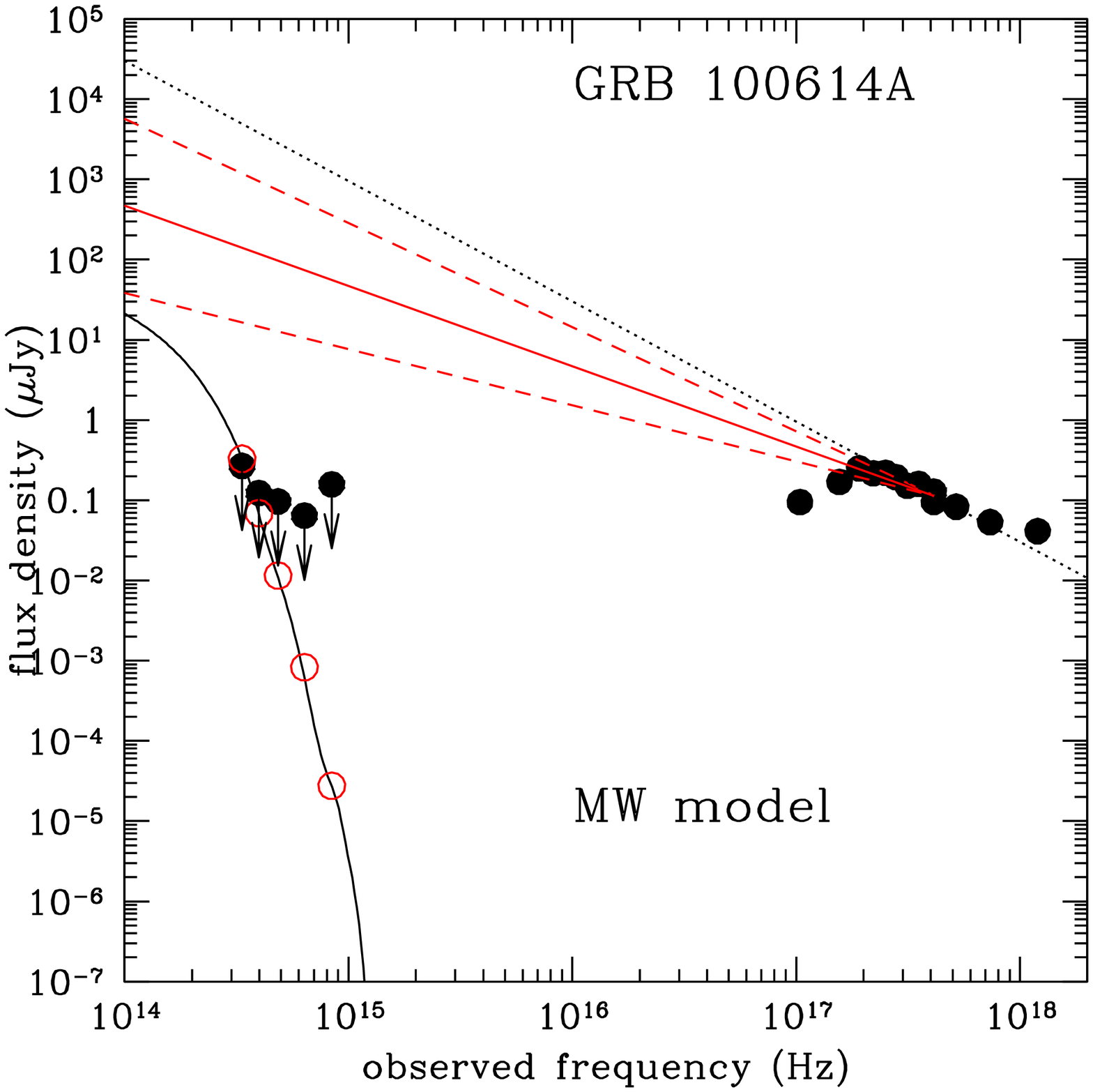}
\includegraphics[angle=-0,width=4.4cm]{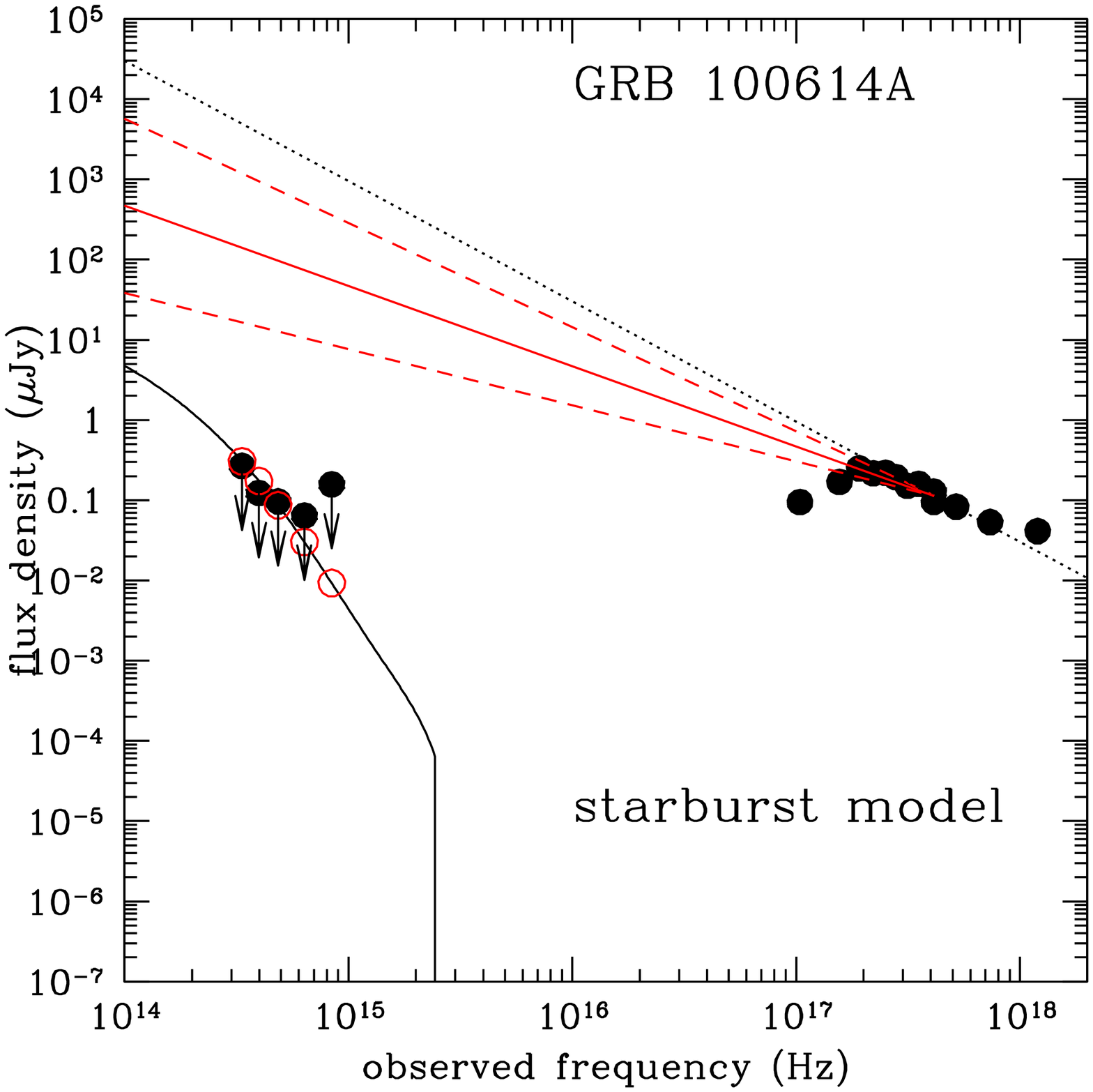}

\includegraphics[angle=-0,width=4.4cm]{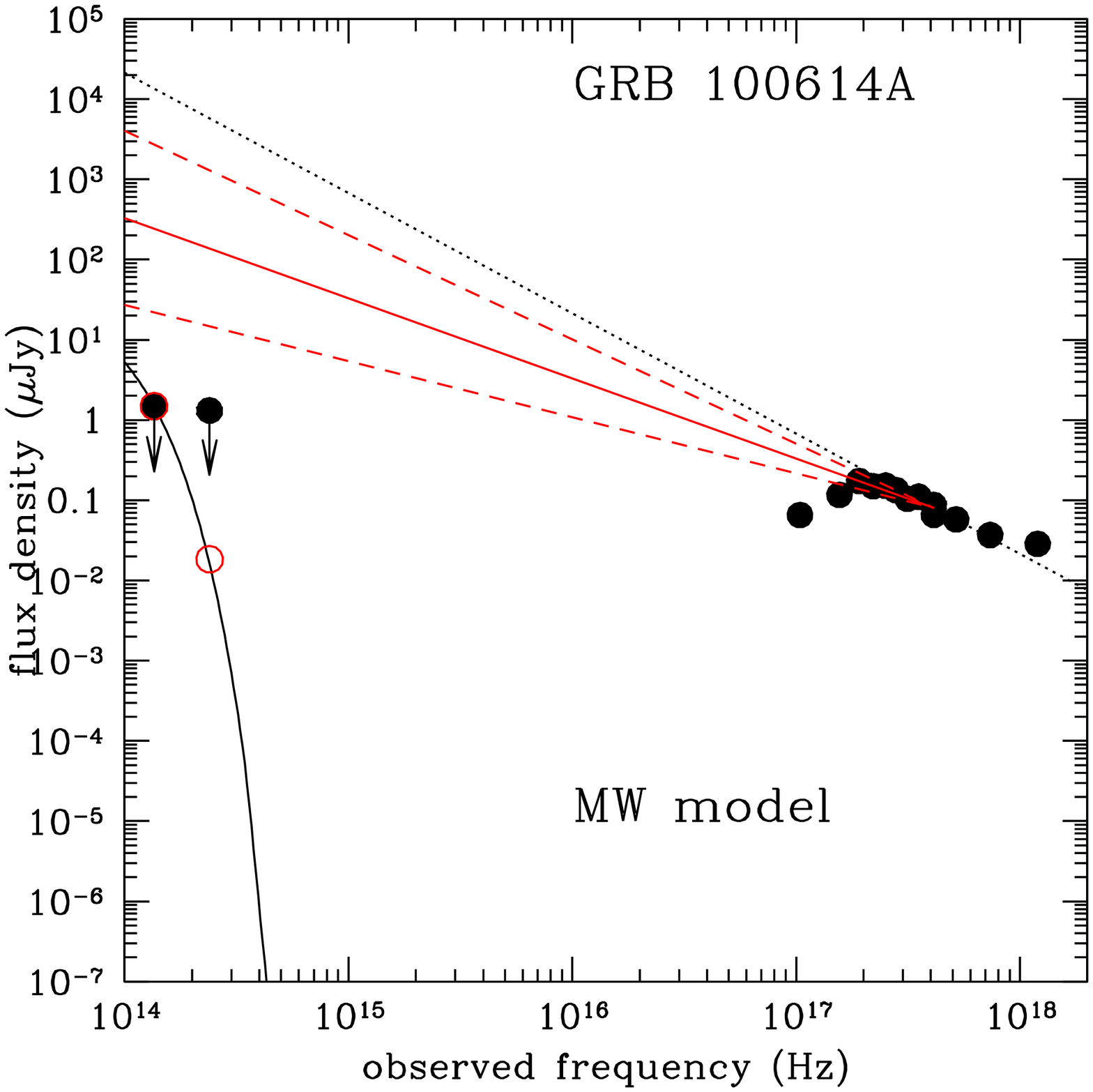}
\includegraphics[angle=-0,width=4.4cm]{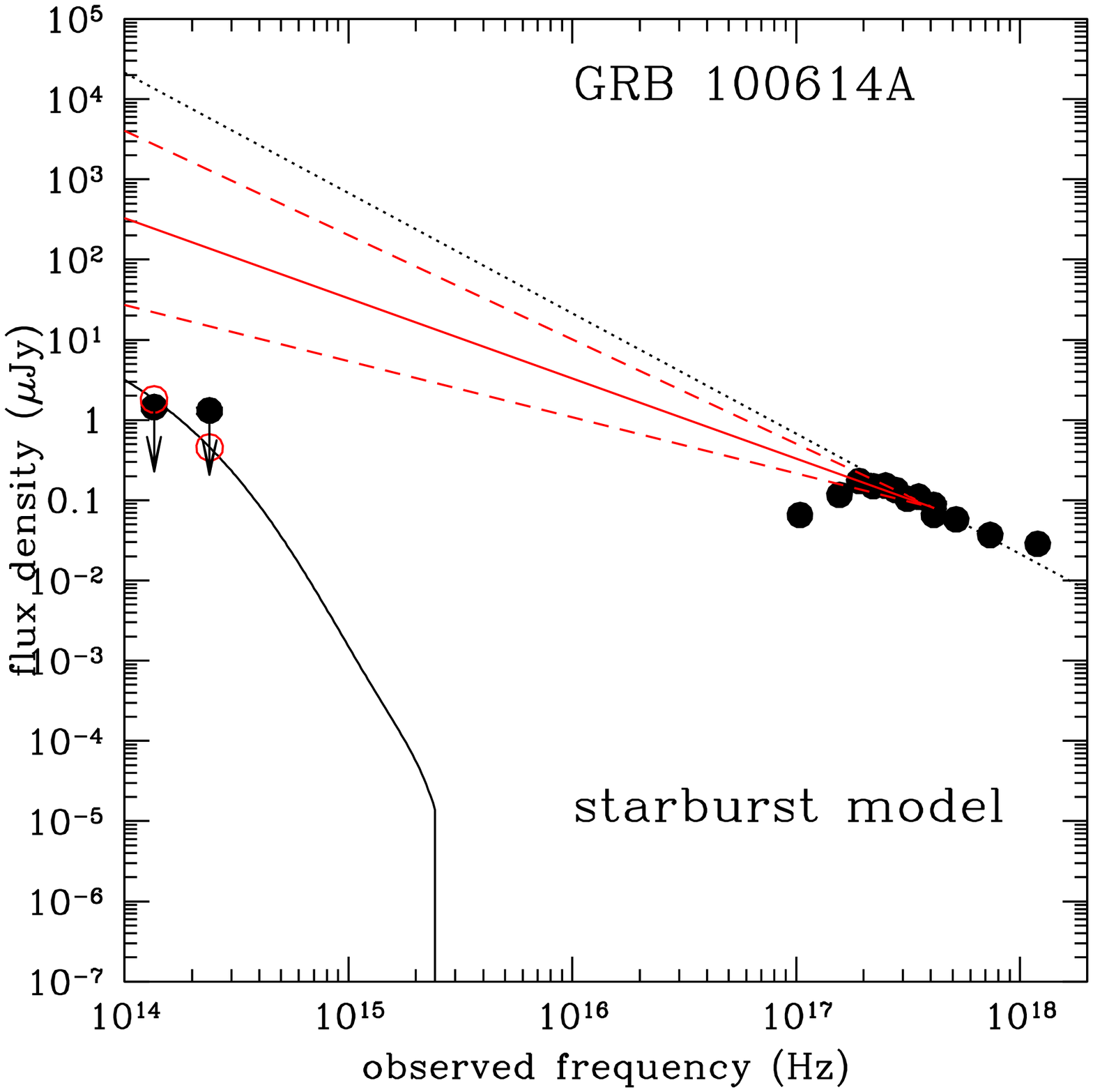}
\caption{GRB\,100614A SEDs at two different epochs: GTC $ugriz$ upper
  limits taken 4.3 hours after the trigger (upper panels) and the NIR
  J and K band taken 9.5 hours after the trigger (lower panels), with
  $\beta_{OX}$ fixed at $\beta_X-0.5$ (solid red line) and the optical
  suppression modelled with the MW or starburst extinction
  curves. Black dotted line represents the best-fit value for the
  X-ray spectral slope ($\beta_X=1.50$). Dashed red lines enclose the
  $90\%$ uncertainty in $\beta_{OX}$. }
\label{spe1}
\end{figure}

 \subsection{GRB\,100615A}

 For this burst, the X-ray steep phase ends 185 s after the trigger,
 and the flat one about 5 ks later. The reddest observations for the
 GRB\,100615A afterglow were performed in the K band 20 minutes
 (Nicuesa et al. 2010), 5.5 hours (Cenko et al. 2010a), and 9.2 days
 (Perley et al. 2010) after the BAT trigger (see Table 1). The first
 observation falls during the plateau phase, while the last two take
 place during the normal phase. The magnitudes of the first
 observation are obtained using the GROND zero point and the 2MASS
 catalog, and are at the $3\sigma$ confidence level (Nicuesa et
 al. 2010). The data of the second observation are calibrated against
 three 2MASS stars and the upper limits are at the $5 \sigma$
 level. The RMS spread when comparing these data with calibrators is
 0.2 in the K band (Cenko et al. 2010a, Cenko, private communication).
 Finally, the data of the last observation are calibrated using two
 2MASS stars and the upper limits are at the $3 \sigma$ level. The
 calibration is accurate to $0.20-0.25$ mag (Perley et al. 2010,
 Perley, private communication). All these magnitudes were corrected
 for Galactic reddening in the direction of this burst. The optical
 upper limits corresponding to the magnitudes of the first observation
 are the following (from the g' to the K band): $f_{g'}<1.0
 ,f_{r'}<1.0$, $f_{i'}<2.3$, $f_{z'}<2.8$, $f_{J}<9.56$, $f_{H}<18.7$,
 $f_{K}<27.9$ $\mu$Jy. The optical upper limits corresponding to the K
 data points of the second and third observations are $f_K<16.1$
 $\mu$Jy and $f_K<4.9$ $\mu$Jy, respectively.

 The 0.3-10 keV energy spectrum was integrated during the plateau, the
 normal and the normal+plateau phases. In all cases, it could be
 closely fitted with an absorbed power law model. The photon indices
 and $N_H$ values are consistent in the three integration intervals,
 thus we assume as the most accurate estimates of the spectral
 parameters, those for the overall integration, which are
 $\Gamma=2.35\pm0.15$ and $N_H=(1.05\pm0.12)\times10^{22}$ cm$^{-2}$
 beyond the Galactic value at the position of the X-ray afterglow
 ($NH_{Gal}=3.3\times10^{20}$ cm$^{-2}$, Kalberla et al. 2005), with a
 count rate to unabsorbed flux conversion factor of
 $1.5\times10^{-10}$ erg cm$^{-2}$ cts$^{-1}$.

 As for GRB\,100614A, we assume this spectral shape for the broad-band
 SED analysis, computing the X-ray normalization level from the light
 curve at the give epochs. The count rate at the selected epochs is
 $c_1=1.0$ cts/s, $c_2=0.2$ counts s$^{-1}$, and $c_3=0.005$ counts
 s$^{-1}$. Converting these values into unabsorbed flux using the
 conversion estimated factor, we find $f_1=1.5\times10^{-10}$ erg
 cm$^{-2}$ s$^{-1}$, $f_2=3.0\times10^{-11}$ erg cm$^{-2}$ s$^{-1}$,
 and $f_3=7.5 \times10^{-13}$ erg cm$^{-2}$ s$^{-1}$. The flux densities
 at 1.7 keV (logarithmic mean of the XRT energy range) are therefore
 $f_{\nu_1}=9.84$, $f_{\nu_2}=1.97$, and $f_{\nu_3}=0.013$ $\mu$Jy.

\begin{figure}
\centering
\includegraphics[angle=-0,width=4.4cm]{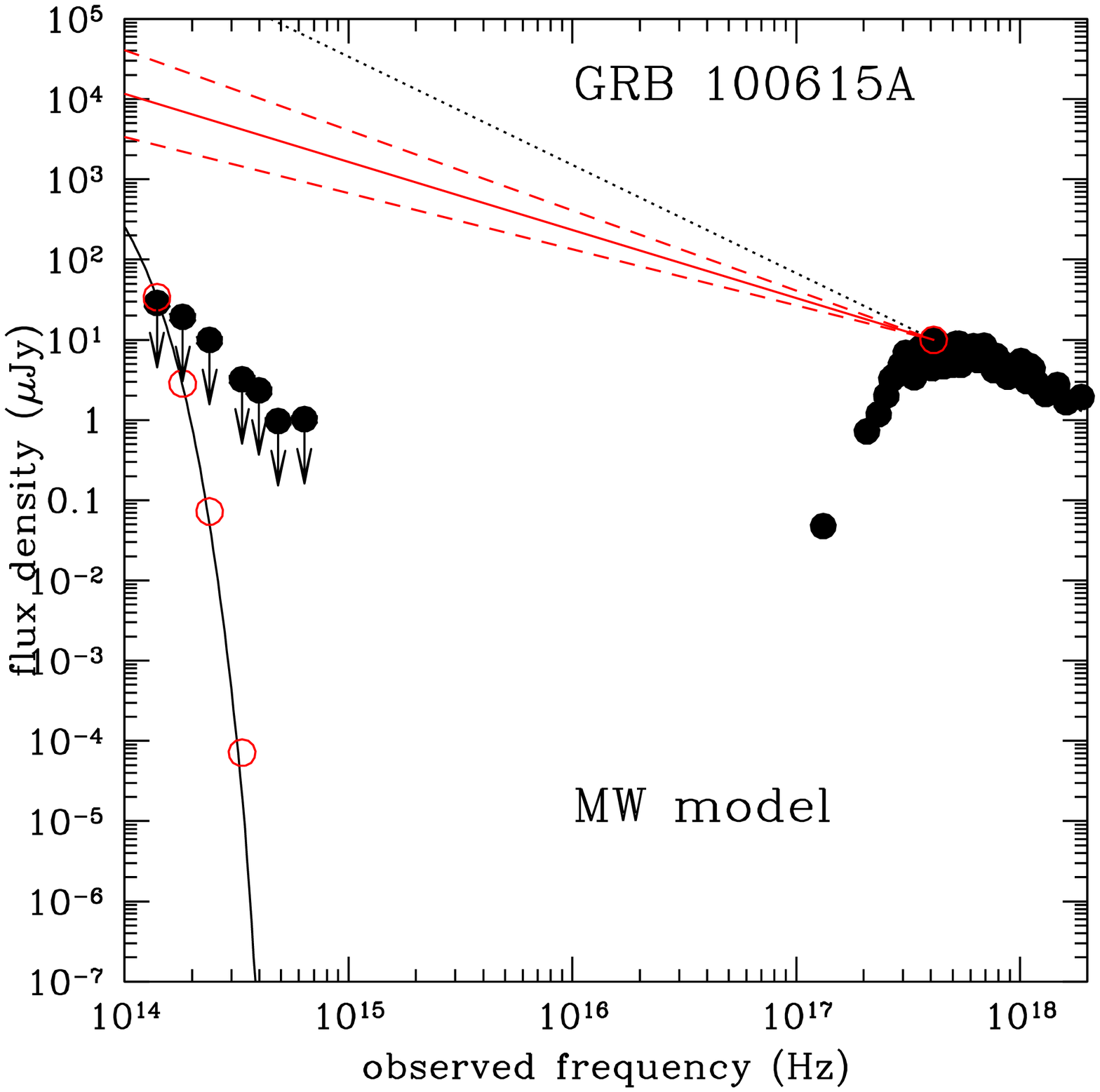}
\includegraphics[angle=-0,width=4.4cm]{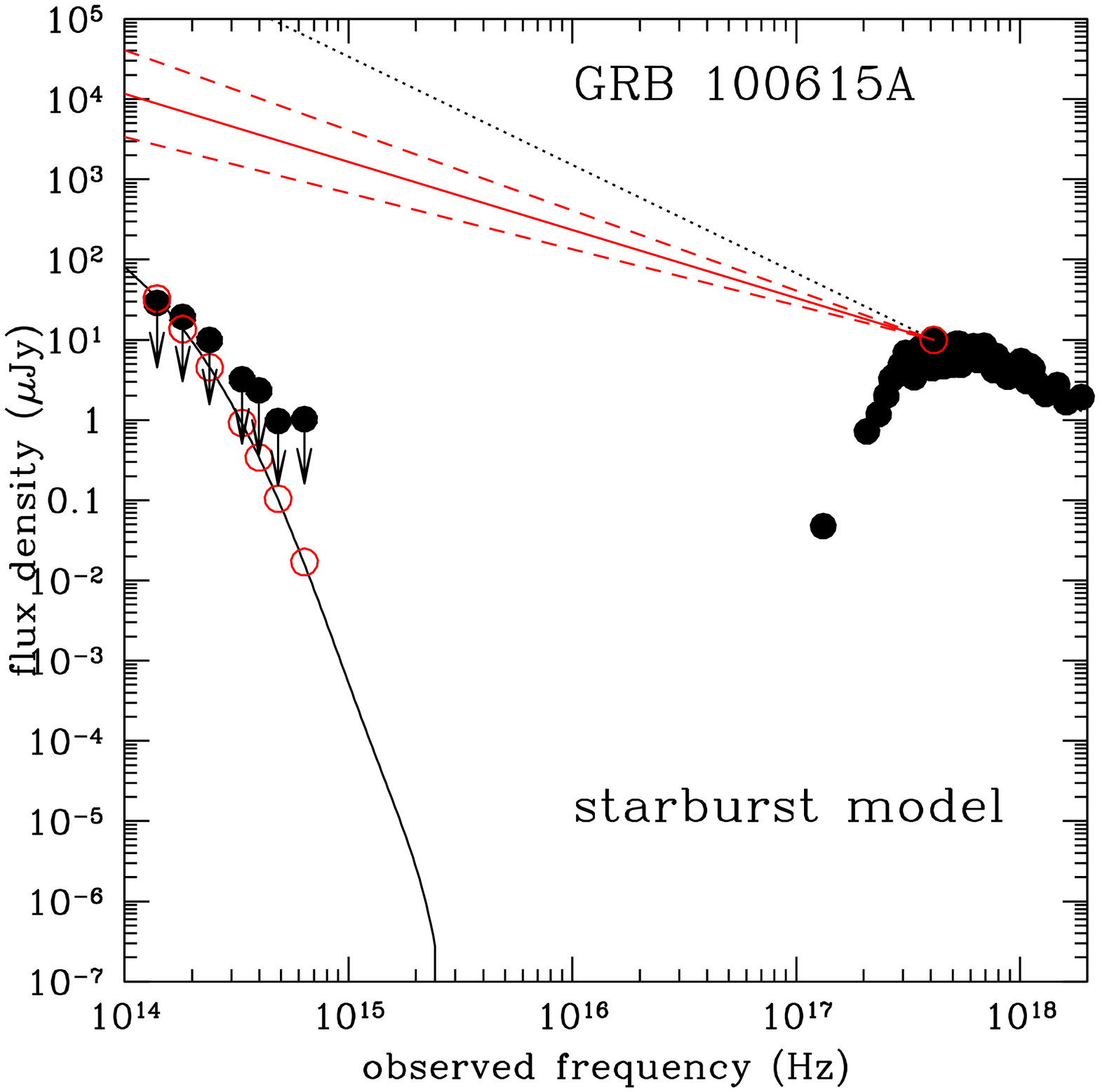}

\includegraphics[angle=-0,width=4.4cm]{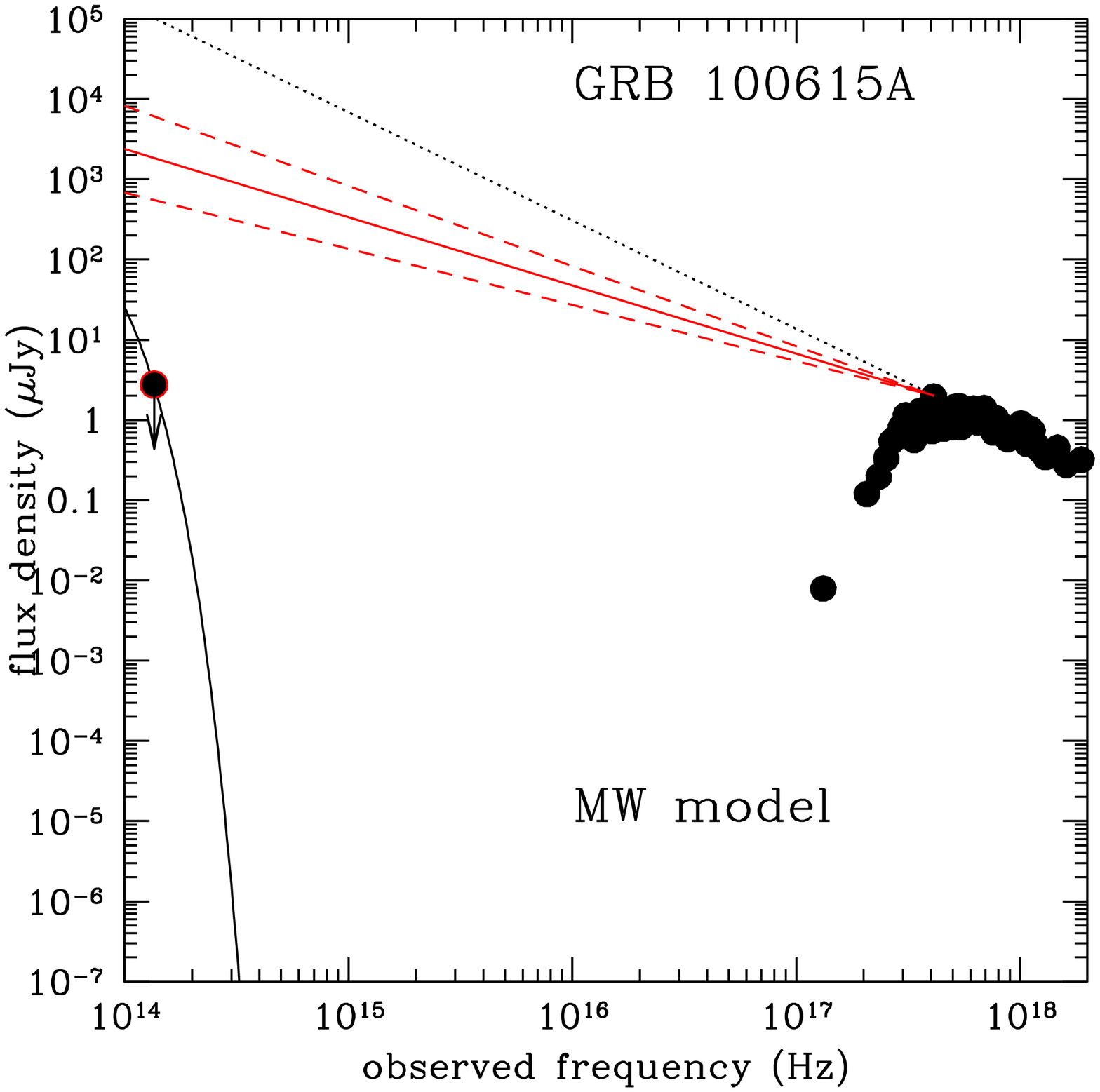}
\includegraphics[angle=-0,width=4.4cm]{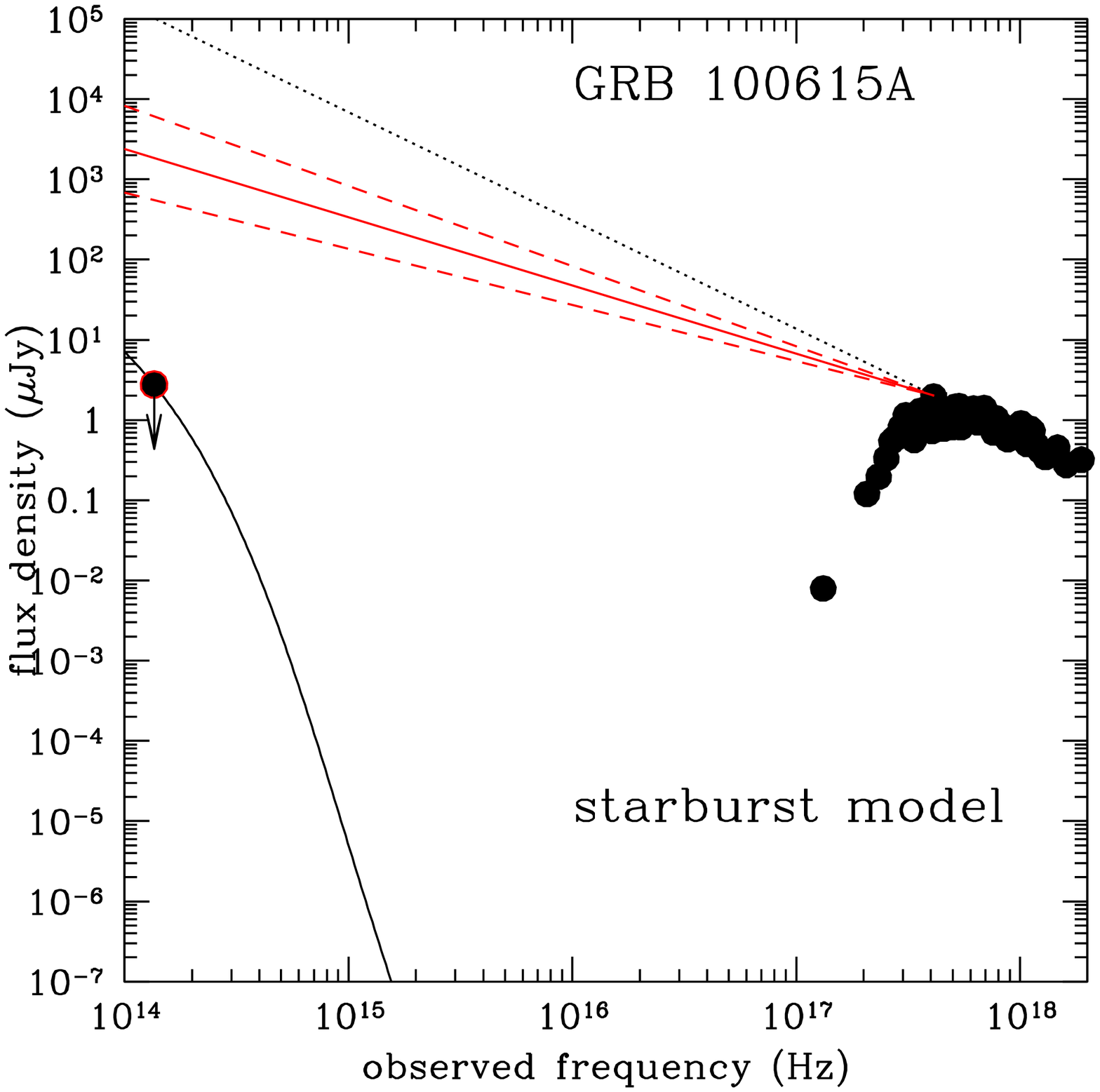}

\includegraphics[angle=-0,width=4.4cm]{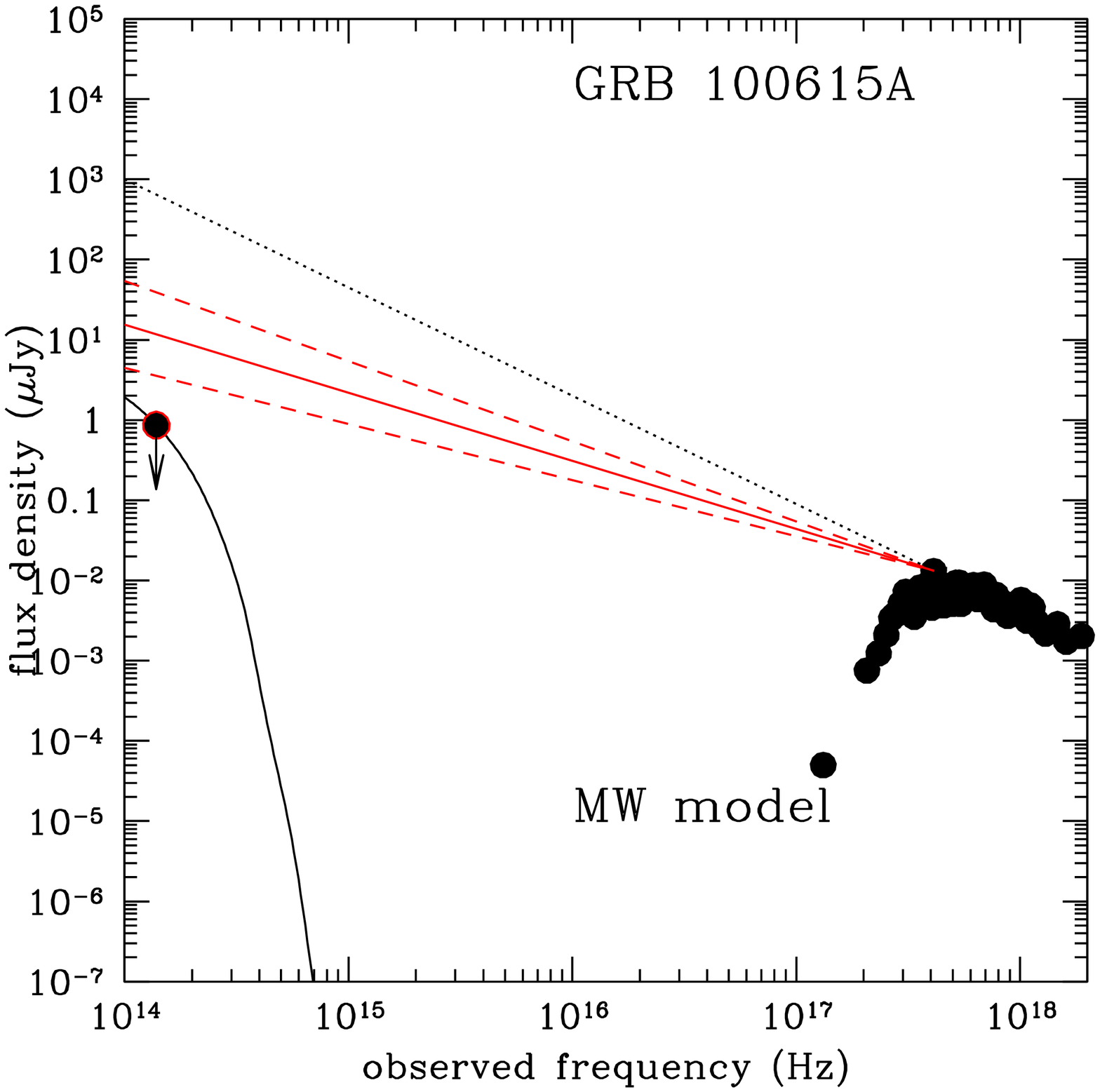}
\includegraphics[angle=-0,width=4.4cm]{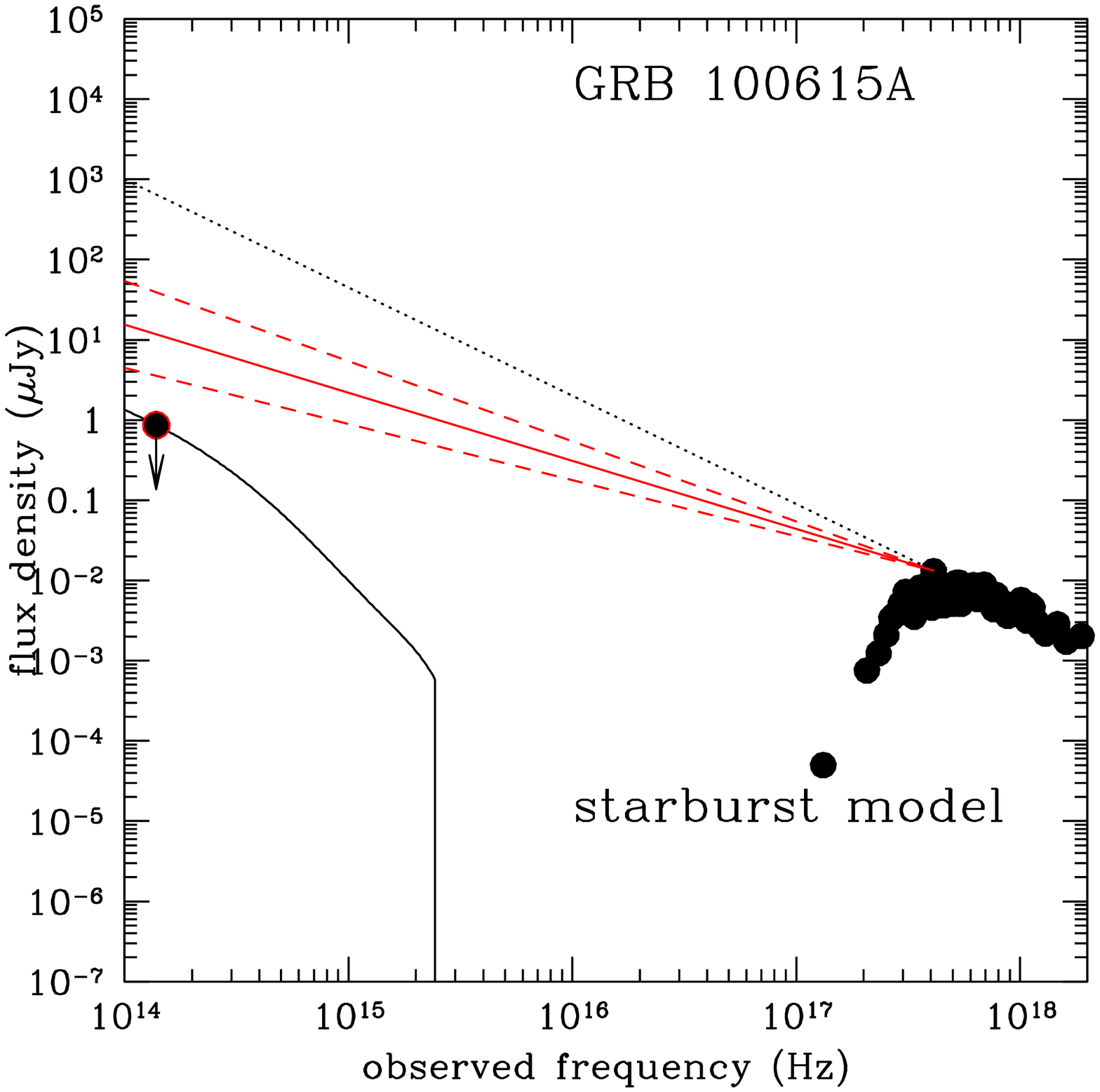}

\caption{GRB\,100615A SEDs at three different epochs: GROND $grizJHK$
  upper limits taken 24 minutes after the trigger (upper panels) and
  Gemini-North K and Ks upper limits taken 5.5 hours and 9.2 days
  after the trigger, respectively (middle and lower panels), with
  $\beta_{OX}$ fixed at $\beta_X-0.5$ and the optical suppression
  modelled with the MW or starburst extinction curves. Black dotted
  line represents the best-fit value for the X-ray spectral slope
  ($\beta_X=1.35$). Dashed red lines enclose the $90\%$ uncertainty in
  $\beta_{OX}$.}
\end{figure}

\begin{table}[ht]
\begin{center}
  \caption{Lower limits to the visual extinction towards GRB\,100614A and GRB\,100615A.} 
{\footnotesize \smallskip
\begin{tabular}{|lccc|ccc|}
  \hline 
  Source             &$t-T$  & $\beta_X$  & $\beta_{OX}$ & $A_{V,MW}$ & $A_{V,SMC}$    & $A_{V,SB}$    \\
  		     & min   &            &             &  mag      & mag            & mag          \\
  \hline
  GRB\,100614A       & $258$         & $1.20$    &              &$8$         & $8$     & $6$        \\
  {\bf GRB\,100614A} & {\bf 258}     & {\bf 1.50}&$<0.1$       &{\bf 13}    & {\bf 13}& {\bf 8}     \\
  GRB\,100614A       & $258$         & $1.80$    &              &$18$        & $18$    & $11$       \\
  GRB\,100614A       & $570$         & $1.20$    &              &$25$        & $31$    & $6$        \\
  {\bf GRB\,100614A} & {\bf 570}     & {\bf 1.50}&$<0.4$       &{\bf 47}    & {\bf 58}& {\bf 11}    \\
  GRB\,100614A       & $570$         & $1.80$    &              &$69$        & $85$    & $17$       \\
  \hline                                            
  GRB\,100615A       & $24$          & $1.20$    &              &$39$        & $47$    & $10$       \\
{\bf   GRB\,100615A} & {\bf 24}      &{\bf 1.35} &$<0.1$       &{\bf 49}    & {\bf 60}& {\bf 13}    \\
  GRB\,100615A       & $24$          & $1.50$    &              &$59$        & $73$    & $16$       \\
  GRB\,100615A       & $330$         & $1.20$    &              &$47$        & $58$    & $12$       \\
{\bf  GRB\,100615A}  & {\bf330}      & {\bf 1.35}&$<0.0$       &{\bf 58}    & {\bf 72}& {\bf 15}    \\
  GRB\,100615A       & $330$         & $1.50$    &              &$69$        & $86$    & $18$       \\
  GRB\,100615A       & $13200$       & $1.20$    &              &$12$        & $15$    & $3$        \\
{\bf  GRB\,100615A}  & {\bf13200}& {\bf 1.35}    &$<0.5$       &{\bf 22}    & {\bf 27}& {\bf 6}     \\
  GRB\,100615A       & $13200$       & $1.50$    &              &$33$        & $40$    & $9$        \\
  \hline
\end{tabular}
}
\end{center}
\end{table}

\section{Results}

We find that GRB fluxes computed from both optical and NIR data are
well below the most conservative extrapolation from X-rays and require
strong absorption using the data taken within 1 day after the
trigger. For GRB\,100615A, very late-time data are available (9.2 days
after the trigger), for which the NIR flux is still below the X-ray
extrapolation.

The GRB\,100614A SED with the reddest flux upper limit (i.e. 570
minutes after the trigger) requires a rest-frame V-band dust
extinction of $A_V\ge47$, $58$, and $11$ mag, assuming a MW, SMC, or a
starburst extinction curve, respectively, and using the best-fit value
for $\beta_X$ (Fig. 2).  Even fixing the optical to X-ray energy
spectral index to its lowest allowed value (within its $90\%$
confidence range), that is, in the most conservative case, results
still provide very high $A_V$ lower limits (Table 2). Less stringent
constraints on $A_V$ are obtained using the GTC $ugriz$ flux upper
limits obtained 258 minutes after the trigger. We obtain $A_V\ge13$
mag with either the MW and SMC extinction curve and $A_V\ge8$ mag with
the starburst case.

For GRB\,100615A, we obtain even tighter lower limits. The SED
extracted 24 minutes after the burst requires $A_V\ge64$, $79$, or
$16$ mag assuming either a MW or SMC extinction curve, or a starburst
attenuation curve, respectively.  These lower limits are still very
high for the SED extracted $5.5$ hours post burst: $A_V\ge58$, $72$,
and $15$ mag for the three extinction recipes. Less critical but still
high values, are obtained even $9.2$ days from the burst: $A_V\ge22$,
$27$, and $6$ mag (Fig.3).  The latter lower limits are lower than the
ones obtained at earlier epochs (but still extreme), possibly due to a
selection effect, since at later times the X-ray flux decreases, but
the optical/NIR upper limits can not become fainter consistently,
owing to the instrument detection limits.

Table 2 reports all the $A_V$ lower limits evaluated from the
GRB\,100614A and GRB\,100615A data, for MW, SMC, and starburst
extinction laws at the given mean observation epochs. These lower
limits are computed for the reported ranges of the X-ray spectral
index ($\beta_X=\Gamma-1$, estimated in Sect. 3) that corresponds to
the minimum, maximum, and mean value of the estimated $90\%$
confidence range. Bold face characters indicate the results obtained
with the most probable X-ray spectral index. Upper limits to the
optical-to-X-ray spectral indices $\beta_{OX}$ are also shown.


\section{Discussion}

We have analyzed two {\it Swift} `dark' GRBs, namely, GRB\,100614A,
and GRB\,100615A. These GRBs are dark according to every definition
proposed until now. They are not detected in the optical/NIR down to
very faint limits, despite follow-up campaigns at ground-based
facilities began a few minutes from the BAT triggers (see Table 1). In
addition, their optical-to-X-ray spectral indices satisfy $\beta_{OX}
< 0.5$ (Jakobsson et al. 2004 criterion) and $\beta_{OX}<\beta_X-0.5$
(van der Horst et al. 2009 criterion).  The identification of these
two GRBs as dark bursts according to the above-mentioned methods, is
the consequence of their intense X-ray flux coupled to the optical/NIR
missing detections. GRB\,100614A (GRB\,100615A) indeed falls in the
upper $30\%$ ($5\%$) of the distribution of {\it Swift}/XRT GRB fluxes
at $11$ hr from the burst (Gehrels et al. 2008).

The outcome of our analysis is surprising. To explain the deepest NIR
upper limits (i.e. the less affected by dust) in terms of a flux
suppression described by either a MW or SMC dust extinction laws, $A_V
> 47$ ($A_V > 58$) mag is needed for GRB\,100614A (GRB\,100615A)
before one day and $A_V>22$ at 9 days after the trigger for
GRB\,100615A. Such extreme $A_V$ values have never been observed
before and require an explanation.
 
As an example, Perley et al. (2009) studied a sample of $29$ {\it
  Swift} bursts rapidly observed by the Palomar $60$-inch telescope
($14$ of which were classified as dark) from April 2005 to March
2008. From both optical to X-ray afterglow spectral analyses and host
galaxy studies (applying the $\beta_{OX}<0.5$ criterion by Jakobsonn
et al. 2004 and assuming in general a SMC extinction law), they find
that more than half of the dark sample is extincted by dust, with
three events featuring $A_V>2-6$ mag, two GRBs possibly being high
redshift events ($4.5<z<7$), and three being underluminous events (not
detected in the optical bands, but not dark according to Jakobsonn et
al. (2004) criterion).  The dust extinction in their sample is
compatible to that observed along the dustiest MW
sightlines. Extinction values of A$_V\le 5$ are found along more than
$700$ Galactic sightlines (Diplas \& Savage 1994). In a handful of
extreme cases, A$_V$ can be higher, up to $\sim 20$ (Predehl \&
Schmitt 1995).  Rol et al. (2007) studied what they called a
`prototype' dark GRB, namely, GRB\,051022. This burst misses any
optical/NIR detection, such as GRB\,100614A and GRB\,100615A, and is
the darkest burst in the paper by van der Horst et al. (2009)
according to their classification criterion. Applying our analysis to
GRB\,051022, we find $A_V > 16$ and $20$ mag assuming a MW and SMC
extinction curve, respectively.  Again, these values are well below
the $A_V$ lower limits we need to fit the deepest NIR upper limits of
GRB\,100615A and GRB\,100614A, using the same extinction curves.

While a SMC-like extinction curve can adequately fit a large fraction
of the dust extinction from GRB host ISM, our present picture of GRB
host galaxies makes an extremely obscured environment of this kind a
very unlikely possibility. Indeed, GRBs, even reddened ones, are
hosted by blue or normal galaxies, which are commonly detected in
ordinary galaxy surveys. This favours a scenario of a host morphology
where the line of sight to the GRB is dusty, i.e., dust obscures only
localized regions (see e.g., Perley et al. 2009 and references
therein). An in situ obscuration appears to be insufficient to be
responsible for the extreme extinction levels we measure.

All these considerations hold if the optical radiation and X-rays are
part of the same synchrotron spectrum. They could originate from
different emission processes or even be produced in different,
independent emission regions. This is possible in particular during
the so-called "shallow phase", where X-ray emission may be dominated
by an emission component that differs from the one from which the
optical flux originates (e.g. Zhang et al. 2006 for a review).
Interesting hints about the early optical to X-ray afterglow spectral
behavior came from Greiner et al. (2011). These authors addressed the
darkness problem by studying all the GRBs observed by the GROND imager
mounted on the 2.2 m MPI/ESO telescope at La Silla (Greiner et
al. 2008).  Broad-band SEDs of those GRBs with simultaneous optical
and X-ray data within the first 240 minutes and after the early X-ray
steep decay (43 GRBs, 39 of which are long) have been extracted and a
simple or a broken power-law spectral model has been fitted to the
data according to the darkness criterion of van der Horst et
al. (2009). The optical to X-ray SEDs of those GRBs at known redshifts
clearly show that in no case the optical fluxes are above the X-ray
extrapolation, a signature that would have confirmed the possible
distinct origin of the early X-ray emission from the optical one. In
particular, modelling the optical suppression from the X-ray
extrapolation as SMC- or MW-type dust extinction, Greiner et
al. (2011) found in all cases a good agreement between the expected
fireball spectral model and the data, and a fraction of $25\% - 40\%$
of the bursts of their sample were found to be dark (different
percentages depend on the definition used), where their darkness can
be explained by moderate extinction ($0.5<A_V<1.5$), or high redshift
($z>5$) for $22\%$ of the dark bursts. These results show the lack of
an evident inconsistency between the optical and X-ray early afterglow
SED. However, the brightness of the X-ray flux may be biased by the
possible presence of a dominant component that differs from the one
responsible for the optical emission: excess of the X-ray to optical
flux ratio could thus mimic a stronger optical absorption than the
real one. Since the two SEDs of GRB\,100614A, the first one of
GRB\,100615A, and (marginally) the second one of GRB\,100615A are all
extracted during the shallow phases of these GRBs, a different origin
of the optical and X-ray emission in these epochs could at least in
part explain the optical darkness of our GRBs. However, we note that
for GRB\,100615A we have extracted a SED at a very late time, about
nine days after the end of the plateau phase, and we have still
obtained very high $A_V$ lower limits ($A_v>20$) assuming either a MW
or a SMC extinction curve. These values are less extreme than that
obtained using the other SEDs, but still very high, suggesting another
or at least a concurring mechanism to account for the GRB darkness.

This complementary explanation could be that local extinction recipes,
such as MW or SMC ones are inadequate for reproducing the optical
suppression in the host galaxies of these two GRBs.  Indeed, the dust
and extinction properties of GRB host galaxies are still poorly known.
For example, modelling the dust absorption using greyer extinction
laws, such as the attenuation curve obtained from the observations of
starburst galaxies proposed by Calzetti (1994), brings GRB\,100614A
and GRB\,100615A to require $A_V$ lower limits that are less extreme,
despite still being very high. More gray extinction curves have
already been invoked for GRB environments by several authors
(e.g. Perley et al. 2009; Stratta et al. 2004, 2005; Chen et al. 2006;
Li et al. 2008).  The shape of the extinction curve provides
information on the dust properties. The study of dust in GRB
environments is particularly useful for studying any evolution of dust
properties and dust production mechanisms up to very high redshifts
and in "normal" galaxies. The latters are indeed more representative
of the majority of the galaxy population, rather than the more extreme
galaxies, such as those hosting quasars.

A mixture of moderate-to-high redshift and extinction can reduce the
dust level necessary to explain the SEDs. In fig. 4, we plot as an
example the $A_V$ values as a function of $z$ obtained for the second
epoch of GRB\,100615A observations. Although the visual extinction is
considerably lower than for the $z=0$ case, $A_V \sim 10$ at $z=2$ and
$A_V \sim 4-5$ at $z=5$ is still required, regardless of the adopted
extinction recipe. Kann et al. (2010) reported the computed rest-frame
$A_V$ values for a vast sample of GRBs with known redshift (see their
fig. 3). All GRBs in that sample are modelled with $A_V \le
1.3$. Similar results are obtained by Greiner et al. (2011) whose SEDs
of GBRs at known redshift are modelled with $A_V \le 1.5$.  We stress
that GRB samples with known redshift are usually biased towards
non-absorbed GRBs, and the comparison between their $A_V$ and that of
our sources must then be taken with caution. However, the Greiner et
al (2011) sample is more than 90\% complete, since only 3 out of the
39 long GRBs observed by GROND miss a redshift estimation. Although
our $A_V$ lower limits in this scenario are well above those estimated
for the Kann et al. (2010) and Greiner et al. (2011) samples, they are
not extremely unlikely. Thus, an intrinsic origin and/or dust
extinction, coupled to a moderately high redshift could explain the
darkness of GRB\,100614A and 15A.

\begin{figure}
\centering
\includegraphics[angle=-0,width=9cm]{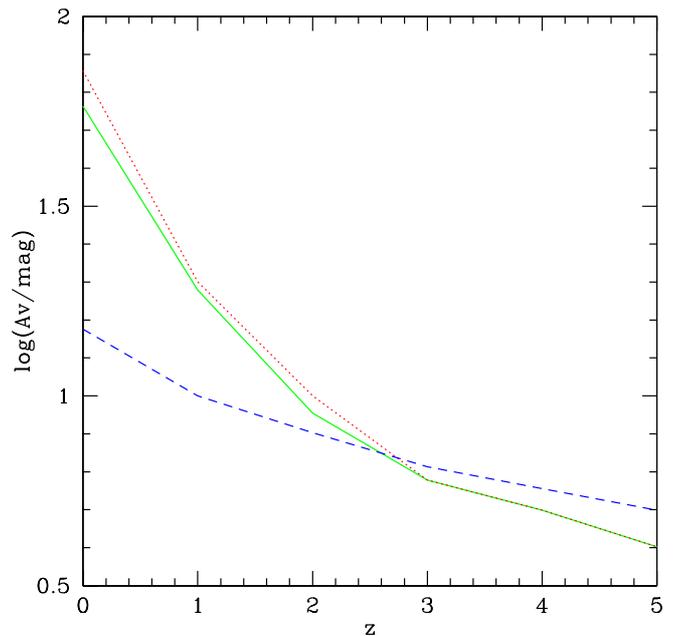}
\caption{The visual extinction $A_V$ as a function of redshift for the
  second epoch SED of GRB\,100615A. Green solid, red dotted, and blue
  dashed lines are for the Milky Way, Small Magellanic Cloud, and
  starburst extinction recipes, respectively.}
\label{spe1}
\end{figure}

A more exotic but intriguing possibility would be that these GRBs are
extremely high redshift events.  Assuming that the lack of any
detection in the reddest NIR band (K-band) is due to Ly$\alpha$
absorption from the intergalactic hydrogen neutral fraction, we can
set a redshift lower limit of $z>17$. The first population of very
massive stars (PopIII, $10^2M_\odot<M<10^3M_\odot$) is expected to
form at $z\sim 20$. Their death is supposed to leave behind black
holes of several tens of solar masses, which could be the early
progenitors of active galactic nuclei. These fast-spinning black holes
have a rotational energy of $\sim 10^{55}$ erg or more that can power
a GRB explosion (see e.g., Komissarov \& Barkov 2010; M\'esz\'aros \&
Rees 2010; Suwa \& Yoka 2011). The isotropic energies of GRB\,100614A
and GRB\,100615A can be estimated from their BAT fluence (Sakamoto et
al. 2010; Palmer et al. 2010). The resulting values assuming $z=18$
are $1.3 \times 10^{54}$ erg and $7.2 \times 10^{53}$ erg,
respectively. Such energies fall in the bright tail of the GRB
distribution, the most energetic burst detected to date, GRB\,080916C,
having $E_{iso}=9\times 10^{54}$ erg (Abdo et al. 2009).  The Amati
relation (Amati et al. 2008) allows for a $2\sigma$ scattering of the
prompt emission energy peak of $0.5-5$ MeV in the rest frame, assuming
an isotropic energy of $\sim 10^{54}$ erg. The reported isotropic
energies must be considered as lower limits both because of the
conservative choice of the $z$ used and because the BAT detector does
not constrain the position of the peak emission preventing a
bolometric estimate of the emitted energy. Thus, that these peaks are
not required in the 15-150 keV spectral fit of the GRB\,100614A and
GRB\,100615A BAT data (Sakamoto et al. 2010; Palmer et al. 2010) is
not surprising.  In addition, according to some models (see e.g., Suwa
\& Yoka 2011), a considerable fraction of the available energy to
produce the GRB must be used to pierce the envelope of the very
massive star. The lack of detection of any host galaxy candidate for
these GRBs represents additional support of this scenario.  On the
other hand, against the high redshift interpretation there is that the
expected GRB rate at $z>17$ is extremely low, between $0.5$ and $1$
GRB every $10$ yr (Bromm \& Loeb 2006). In addition, there is the
non-negligible equivalent hydrogen column density value measured for
GRB\,100614A and GRB\,100615A from X-ray spectroscopy.  For example,
at $z=18$, the rest-frame column density would be roughly
$(1+z)^{8/3}$ times higher than the measured value, thus
$\sim300-3000\times10^{22}$ cm$^{-2}$ for our GRBs. These high
hydrogen column densities would require an extremely dense environment
surrounding the GRB, which challenges this scenario.

A possible way to differentiate between extremely high-redshift,
exotic extinction recipes and emission from distinct components for
sources such as these, would be to search for the afterglow in the
mid- or far-IR bands. A non-detection also in these bands could hardly
be explained using any extinction law and would definitely rule out a
high redshift origin. A multiband detection compatible with the X-ray
flux at late times (e.g. $> 10^5$ s after the trigger, i.e. after the
end of the "shallow phase") would instead favour the fireball
model. In this case, given the lack of NIR detections, the origin of
the darkness would be in exotic extinction or high redshift, depending
on the spectral shape in the mid- and far-IR bands.

\begin{acknowledgements}
  This work made use of data supplied by the UK Swift Science Data
  Centre at the University of Leicester. We are indebted to S. Guziy,
  S.B. Cenko, and D.A. Perley for providing us with details on the
  analysis performed in their GCNs.
\end{acknowledgements}

\end{document}